\documentclass[10pt, conference, compsocconf]{IEEEtran}

\usepackage{bm}
\usepackage{cite}
\usepackage[dvipdfmx]{graphicx}
\usepackage[cmex10]{amsmath}
\begin{document}

\title{Bursting activity spreading through asymmetric interactions}

\author{\IEEEauthorblockN{Tomokatsu Onaga}
\IEEEauthorblockA{Department of Physics, Kyoto University\\
Kyoto 606-8502, Japan\\
Email: onaga@scphys.kyoto-u.ac.jp}
\and
\IEEEauthorblockN{Shigeru Shinomoto}
\IEEEauthorblockA{Department of Physics, Kyoto University\\
Kyoto 606-8502, Japan\\
Email: shinomoto@scphys.kyoto-u.ac.jp}}

\maketitle

\begin{abstract}
People communicate with those who have the same background or share a common interest by using a social networking service (SNS). News or messages propagate through inhomogeneous connections in an SNS by sharing or facilitating additional comments. Such human activity is known to lead to endogenous bursting in the rate of message occurrences. We analyze a multi-dimensional self-exciting process to reveal dependence of the bursting activity on the topology of connections and the distribution of interaction strength on the connections. We determine the critical conditions for the cases where interaction strength is regulated at either the point of input or output for each person. In the input regulation condition, the network may exhibit bursting with infinitesimal interaction strength, if the dispersion of the degrees diverges as in the scale-free networks. In contrast, in the output regulation condition, the critical value of interaction strength, represented by the average number of events added by a single event, is a constant $1-1/\sqrt{2} \approx 0.3$, independent of the degree dispersion. Thus, the stability in human activity crucially depends on not only the topology of connections but also the manner in which interactions are distributed among the connections.
\end{abstract}
%\IEEEpeerreviewmaketitle

%\begin{IEEEkeywords}
%Complex networks;
%\end{IEEEkeywords}

\section{Introduction}

\begin{figure}[!ht]
\centering
\includegraphics[width=3.4in]{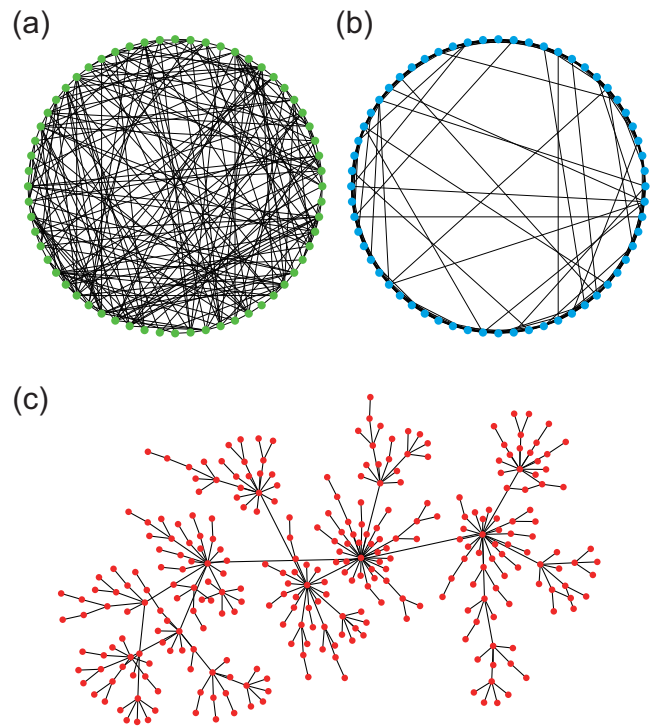}
\caption{Different kinds of network topology. (a) random network (Erd\H{o}s-R\'enyi model); (b) small-world network (Watts and Strogatz model); (c) scale free network (Barab\'asi-Albert model).}
\label{topology}
\end{figure}

In these days, people send messages or share news in a social networking service (SNS). Message occurrences exhibit bursting in their occurrence rate in response to major events such as earthquakes or the world cup results. It is known that bursting may also occur rather spontaneously through word-of-mouth communication, even if there are no triggering major events~\cite{Barabasi05,Sornette04,Crane08,malmgren2008}. The emergence of such spontaneous or endogenous bursting activity may depend not only on the topology of connections among people, such as random, small-world, or scale-free (Fig.~\ref{topology}), but also crucially on the distribution of interaction strength among the connections~\cite{karsai2011,rocha2011,rocha2013,takaguchi2013}. Here, we analyze simple model system to reveal the generic dependence of bursting activity on such network parameters. In the model system, each event is idealized as a point event, by ignoring complex details and simplifying such that every event has equal probability of having influence on other nodes; events are generated stochastically, whereby each event is derived from an underlying rate, allowing spontaneous occurrence; every node is influenced by the events in the connected nodes in a manner such that the underlying rate for generating future events is modified (typically increased). The above-mentioned process may be mathematically formulated as the self-exciting process or the Hawkes process~\cite{Hawkes71a}, which has been widely applied to the analysis of earthquakes~\cite{Ogata88}, genome sequences~\cite{Reynaud10}, neuronal activity~\cite{Krumin10,Rotter13}, urban crime~\cite{Mohler11,White14}, and human activity~\cite{Blundell12,Mitchell10,Masuda13,Filimonov12}.

When applying a network model to the analysis of human activity, the nodes, edges, and events occurring at nodes, respectively correspond to persons, their relation or connections, and messages dispatched from individual persons. Generally, the strength of interaction is not homogeneous over the connections between persons as illustrated in the following.

\begin{figure}[!ht]
\centering
\includegraphics[width=3.4in]{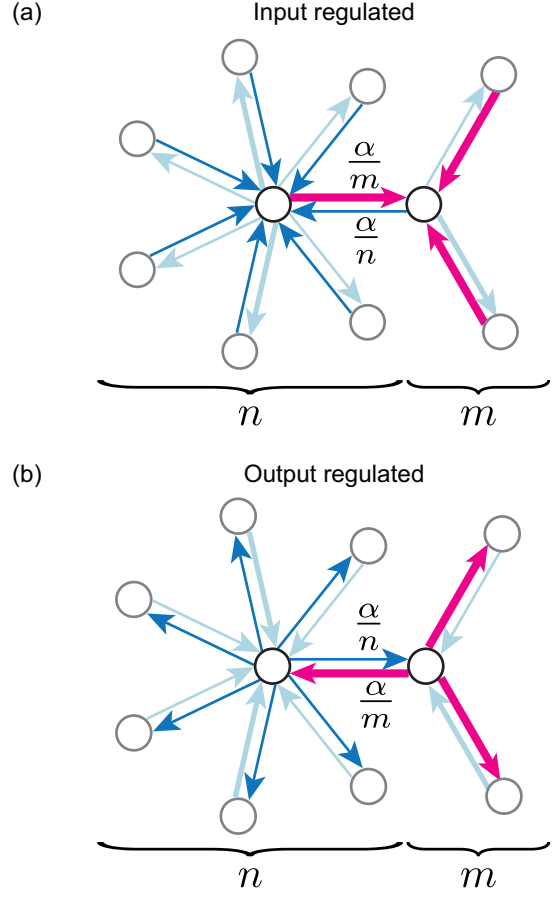}
\caption{Two types of asymmetric interactions. (a) Input regulated condition: Each node is influenced from connected nodes with susceptibility inversely proportional to the degree of the receiver node. (b) Output regulated condition: Every node influences with susceptibility inversely proportional to the degree of the sender node.}
\label{connections}
\end{figure}

In an SNS, the number of friends of each person ranges widely from a few to thousand~\cite{Kwak10,Bakshy11,Cha2012,Ugander11}. The ``popular'' people who have a large number of friends may be more influential than ``unpopular'' people who have fewer friends. Regarding the susceptibility to information, however, there would not be a large difference between individual persons, because a single person has limited capacity for information processing and cannot read and react to all of the received messages. We simplify this situation as ``input regulated,'' such that the input interaction from each sender to the person of interest is scaled inversely proportional to the degree (the number of connections) of the receiver node. In this situation, the interaction is generally not bidirectionally symmetric (Fig.~\ref{connections} (a)).

The distribution of interactions could be different in other situations. When considering the infection caused by diseases, there may not be large differences in the susceptibility between connections; the probability of getting infected from a single-chance encounter would be independent of the total number of people encountered. In this case, the interaction is not regulated by the number of connections, but would be nearly constant.

As a case opposite to the interactions in an SNS, we may consider a process in which every person donates a fixed amount of money to connected people and such donations lead to other donations. In such a case, the amount of donation per acceptor is smaller if a donor is a popular person having a large number of friends. We may idealize that the interaction is ``output regulated,'' such that the output interaction from the person to each receiver is scaled inversely proportional to the degree of the sender or the donor (Fig.~\ref{connections} (b)).

In this study, we analyze the conditions for networks to exhibit endogenous fluctuation in their event occurrence rates by comparing these two extreme conditions, ``input regulated'' and ``output regulated.'' This paper is organized as follows. In section 2, we introduce the self-exciting process, with the simple derivation of average event occurrences. In section 3, we introduce the condition on which the rate fluctuation can be detected from the event series. In section 4, we apply the detectability condition on the self-exciting process to analytically obtain the critical interaction strength above which the system exhibits fluctuation or bursting in the occurrence rate of events. In section 5, we compute the critical conditions of the interaction strength for population activity. In section 6, we apply it to two types of networks with input regulated and output regulated interactions. In section 7, we discuss the results.

\section{Multi-dimensional self-exciting process}

\begin{figure}[!ht]
\centering
\includegraphics[width=3.4in]{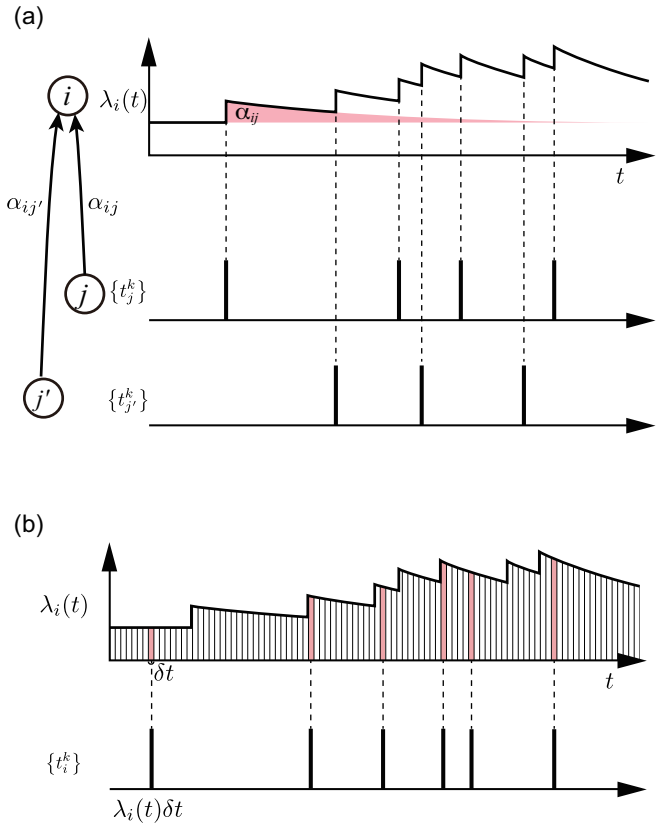}
\caption{Multi-dimensional self-exciting process. (a) Interaction between nodes. The rate of event occurrence of $i$th node, $\lambda_i(t)$, is modulated according to events generated at $j$th and $j'$th nodes. (b) Event generation. Events $\{ t_i^k\}$ are derived randomly from the underlying rate $\lambda_i(t)$.}
\label{figschematic}
\end{figure}

The manner in which persons (nodes) are interacting with messages (events) is described by the multi-dimensional self-exciting process proposed by Hawkes~\cite{Hawkes71a}, which may be represented as,
\begin{equation}
\lambda_i(t)=\rho_i+\sum_{j=1}^N \alpha_{ij} \sum_k f(t-t_j^k),
\label{mdhawkes}
\end{equation}
where $\lambda_i(t)$  represents the rate of event occurrence in the $i$th node at time $t$. $\rho_i$ is the original base rate. $t_j^k$ is the occurrence time of the $k$th event in the $j$th node. $\alpha_{ij}$ is the expected number of events increased in $i$th node by the influence of a single event of $j$th node (Fig.~\ref{figschematic}(a)). $f(t)$ is the kernel function representing the time course of the supplementary probability, satisfying the causality, $f(t) = 0$ for $t < 0$, and the normalization, $\int_0^{\infty}f(t) dt = 1$.

Events are derived randomly in time from an underlying rate according to inhomogeneous Poisson process. This process can be realized by repeating Bernoulli trials for generating events with the probability of $\lambda (t) \delta t$ in every small interval of $\delta t$ (Fig.~\ref{figschematic}(b)).

Given an interaction matrix $\bm{A} \equiv \{ \alpha_{ij} \}$, the average occurrence rates $\bm{\langle \lambda \rangle} \equiv \{\langle \lambda \rangle_i\}$ are obtained from base rates $\bm{\rho} \equiv \{\rho_i\}$ as
\begin{equation}
\bm{\langle \lambda \rangle}=\bm{\rho}+\bm{A} \bm{\langle \lambda \rangle}.
\end{equation}
The system may exhibit pandemic explosion in the event occurrences in the following situations: if interactions make a single event induce more than one events on an average, or equivalently if the largest eigenvalue of the interaction matrix $\bm{A}$ exceeds unity. Otherwise, all nodes keep generating events in finite rates given by~\cite{Rotter13},
\begin{equation}
\bm{\langle \lambda \rangle} = \bm{C} \bm{\rho},
\end{equation}
where $\bm{C}$ represents effective interactions,
\begin{equation}
\bm{C} \equiv \left(\bm{I}-\bm{A}\right)^{-1}.
\label{effectiveinteraction}
\end{equation}

\section{Estimating the rate fluctuation}

In the self-exciting process, the rate for generating events inevitably fluctuates over time due to the influence of generated events, as in Eq.(\ref{mdhawkes}). Given a series of events, the underlying rate may be inferred using rate estimators such as a time histogram. A histogram may be optimized by selecting the bin size so that the mean integrated squared error (MISE) between the underlying rate $\lambda(t)$ and the histogram $\hat{\lambda}(t)$ is minimized. Here the MISE is defined as,
\begin{equation}
S=\lim_{T \to \infty}\frac{1}{T}\int_0^T \left\langle \left(\lambda(t)-\hat \lambda(t)\right)^2 \right\rangle dt,
\label{mise}
\end{equation}
where $T$ is the entire observation interval and the bracket $\left\langle \cdot \right\rangle$ represents the ensemble average over possible realization of the stochastic process. It is possible to determine the optimal bin size $\Delta^*$ solely from a series of events, even if the underlying rate $\lambda(t)$ is not known~\cite{Shimazaki07}. 

However, it may occur that the rate fluctuation is ``drowned out'' by the irregular occurrence of events, if the fluctuation in the underlying rate is small or rapidly fluctuating over time. This can be confirmed by the fact that the optimal bin size for a histogram diverges, because the divergence implies that any histogram of a finite bin size produces a larger error than simply assuming a constant rate. Otherwise, the fluctuating rate can be ``detected'' properly. 

One of the authors obtained the condition under which an optimal bin size diverges or not~\cite{Koyama04,shintani12}. It can be shown (see APPENDIX) that the MISE optimal binsize is finite if the rate fluctuation $\delta\lambda(t) \equiv \lambda(t) - \langle \lambda \rangle$ satisfies the following condition:
\begin{equation}
\int_{-\infty}^{\infty}\langle \delta\lambda(t+s)\delta\lambda(t)\rangle ds > \langle \lambda\rangle,
\label{estimation}
\end{equation}
and otherwise the optimal bin size diverges, implying that the rate fluctuation cannot be estimated properly from a series of events derived from the underlying rate. This detectability condition in Eq.(\ref{estimation}) derived from the MISE optimization of a histogram turned out to be identical to the one derived from the marginal likelihood maximization of the Bayesian rate estimator~\cite{Koyama07}, implying that this may be a universal bound. 

\section{Transition in the self-exciting process}

In the preceding section, we revealed that the condition for estimating the rate fluctuation is given in Eq.(\ref{estimation}) in terms of the autocorrelation of the underlying rate,
\begin{eqnarray}
\phi(s) &\equiv& \langle \delta \lambda(t+s) \delta \lambda (t) \rangle \nonumber \\
&=& \langle \lambda(t+s) \lambda (t) \rangle - \langle \lambda \rangle^2.
\label{phi}
\end{eqnarray}
The Fourier transform of the autocorrelation
\begin{equation}
\tilde\phi_{\omega} \equiv \int_{-\infty}^{\infty} \phi(t) \exp {(-i\omega t)} dt,
\end{equation}
or equivalently the power spectrum of the rate fluctuation was obtained by Hawkes~\cite{Hawkes71a}. For the one-dimensional Hawkes process ($N=1$, $\alpha_{ij}=\alpha$), the power spectrum is obtained as~\cite{Onaga14},
\begin{equation}
\tilde\phi_{\omega} = \left(\frac{1}{(1-\alpha \tilde f_{\omega})(1-\alpha \tilde f_{-\omega})} - 1\right)\langle \lambda \rangle ,
\label{fouriercorrelation}
\end{equation}
where $\tilde f_{\omega}$$ \equiv \int_{-\infty}^{\infty} f(t) \exp {(-i\omega t)} dt$ is the Fourier transform of the kernel function.

Because the condition for detecting fluctuation is given in terms of the autocorrelation of rate fluctuation, Eq.(\ref{estimation}), or
\begin{equation}
\tilde\phi_{0} =\int_{-\infty}^{\infty}\langle \delta\lambda(t+s)\delta\lambda(t)\rangle ds > \langle \lambda\rangle,
\end{equation}
the condition for the self-exciting process to exhibit bursting is obtained in combination with Eq.(\ref{fouriercorrelation}) as
\begin{equation}
\frac{1}{\left(1-\alpha \right)^2} > 2.
\end{equation}
It should be noted that the bursting condition is given independent of the time course of the supplementary probability $f(t)$, because the computation of $\tilde\phi_{0}$ requires only the normalization condition for the kernel, $\tilde f_0=\int_{-\infty}^{\infty} f(t) dt =1$. Thus the rate fluctuation in the one-dimensional self-exciting point process is detectable or inferable (if excitability is larger than critical value) or undetectable (if excitability is smaller than the critical value). The critical value is given by~\cite{Onaga14}
\begin{equation}
\alpha_c=1-1/\sqrt{2} \approx 0.3.
\label{criticalexcitability}
\end{equation}
Note that this bursting transition occurs even if the excitability is much smaller than $\alpha=1$, at which the pandemic explosion of events occurs. 

In our previous study~\cite{Onaga14}, we have extended the theory to multi-dimensional self-exciting processes to discuss the detectability of fluctuating activity in networks given in Eq.(\ref{mdhawkes}). We represent the pairwise correlation of the rate fluctuation by a matrix $\bm{\phi}(s) \equiv \{ \phi_{ij}(s) \}$ given the elements,
\begin{eqnarray}
\phi_{ij}(s) &\equiv& \langle \delta \lambda_i(t+s) \delta \lambda_j (t) \rangle \nonumber \\
&=& \langle \lambda_i(t+s) \lambda_j (t) \rangle - \langle \lambda_i \rangle \langle \lambda_j \rangle.
\label{mdphi}
\end{eqnarray}
Similar to the one-dimensional process, we may obtain the Fourier image of the correlation matrix $\tilde{\bm\phi}_{\omega}$ \cite{Hawkes71b}. In particular, we may obtain the Fourier zero-mode as 
\begin{equation}
\tilde{\bm{\phi}}_0 = \bm{C} \bm{\Lambda}\bm{C}^T-\bm{\Lambda},
\label{mdphi0}
\end{equation}
where $\bm{\Lambda} \equiv {\rm diag}\left(\bm{\langle \lambda \rangle}\right)= {\rm diag} \left(\bm{C} \bm{\rho} \right)$. 

The fluctuating condition for each node in event generation is obtained by applying the inequality given in Eq.(\ref{estimation}) to the element of $\tilde{\bm\phi}_0$. Using the correlation given in Eq.(\ref{mdphi0}), we obtain the condition for $i$th node's activity to fluctuate, as
\begin{equation}
\left(\bm C\bm\Lambda\bm C^T\right)_{ii}>2\bm\Lambda_{ii}=2 \langle \lambda_i \rangle.
\label{conditionith}
\end{equation}

\begin{figure*}[!ht]
\centering
\includegraphics[width=7in]{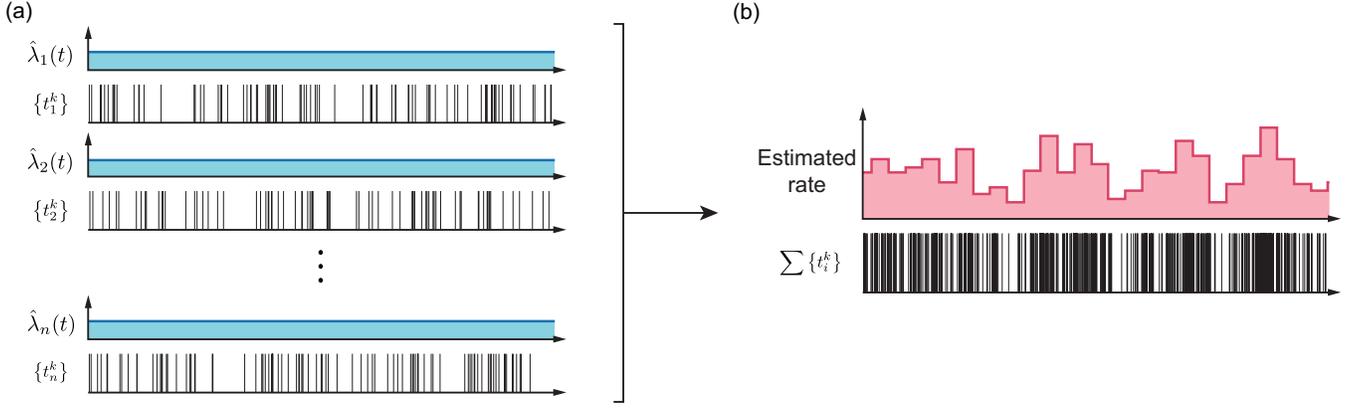}
\caption{Event occurrences in individual nodes and the summed event series. (a) The case in which fluctuations are not detectable from any single nodes. (b) Summed event series may exhibit fluctuation in the occurrence rate.}
\label{figsummed}
\end{figure*}

\section{Bursting transition in population activity}

Even in a situation where fluctuations are not detectable from any single nodes, in which case the inequality given in Eq.(\ref{conditionith}) does not apply for every node, it may occur that fluctuation is estimated from the summed activity of multiple nodes (Fig.~\ref{figsummed}). This is because the signal to noise ratio may increase by superposing noisy data. To see whether this may occur for the activity of the entire population, we examine the estimation condition for the summed rate given as
\begin{equation}
\sum_{i=1}^N \lambda_i(t).
\end{equation}

The correlation of the summed rates are given by
\begin{equation}
\Phi(s) \equiv \sum_{i,j} \langle \delta \lambda_i(t+s) \delta \lambda_j (t) \rangle = \sum_{i,j} \phi_{ij}(s),
\label{PPhi}
\end{equation}
where $\sum_{i,j}$ represents the sum over all element pairs, $\sum_{i=1}^N \sum_{j=1}^N$. Accordingly the integration of correlation is 
\begin{equation}
\tilde{\Phi}_0 = \sum_{i,j} \{\tilde{\bm\phi}_0\}_{ij}.
\label{PPhi0}
\end{equation}
Thus the condition for the summed activity to fluctuate over time is given as
\begin{equation}
\sum_{i,j}\left( \bm C\bm\Lambda\bm C^T \right)_{ij}>2\sum_{i,j}{\bm\Lambda}_{ij}=2\sum_{i} \langle \lambda_i \rangle.
\label{conditionall}
\end{equation}

\section{Fluctuation in two types of asymmetric interactions}

To elucidate the difference in bursting conditions between different types of asymmetric interactions, we simplify the situation such that the mean rates for messaging are identical for all people, $\langle \lambda_i \rangle$ = constant, for $i=1, 2, \cdots, N$. In this situation, the condition for the rate fluctuation given in Eq.(\ref{conditionall}) is 
\begin{eqnarray}
\frac{1}{N}\sum_{i,j}\bm C\bm C^T &=& \frac{1}{N}\sum_i \sum_j \sum_\ell C_{i\ell}  C_{j\ell} \nonumber \\
&=& \frac{1}{N}\sum_\ell \left( \sum_i C_{i\ell} \right)^2 > 2.
\label{condition_for_all}
\end{eqnarray}
Note that we need regulation of the interaction strength to avoid a pandemic explosion, wherein we may obtain the finite effective interaction given in Eq.(\ref{effectiveinteraction}). This may be secured if interactions are regulated at the input or output, as shown in the below.

The critical condition for the above-mentioned transition may be obtained analytically via mean field approximation~\cite{Barrat08}, in which the probability that the node of degree $k_0$ is linked to the node of degree $k$ results in a value independent of $k_0$, given as
\begin{equation}
P(k|k_0)=\frac{k}{\langle k \rangle}P(k),
\end{equation}
where $\langle k \rangle \equiv \sum_k k P(k)$ is the average degree of a node.

\subsection{Input regulation condition}
\begin{figure}[!ht]
\centering
\includegraphics[width=2.6in]{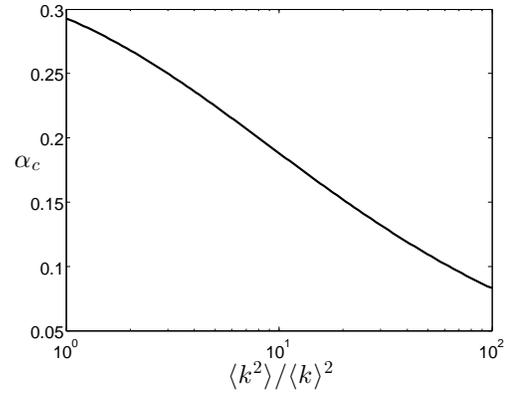}
\caption{The critical interaction $\alpha_c$ depending on the dispersion of the degrees or the numbers of connections $\langle k^2\rangle/\langle k \rangle^2$.}
\label{fig01}
\end{figure}

The condition that the input to each node is regulated is represented in such a way that the influence in each interaction is scaled inversely proportional to the number of input connections,
\begin{equation}
\alpha_{i,\ell} = \alpha / k_i,
\end{equation}
where $k_i$ is the degree of $i$th node (the receiver) and $\alpha$ is a constant representing the entire input that a single node is receiving. For the case that the degree of the sender ($\ell$th node) is $k_\ell$, we compute the summand in Eq.(\ref{condition_for_all}):
\begin{equation}
\sum_i C_{i\ell} = \sum_i (\bm I+\bm A+\bm A^2+\cdots)_{i \ell}.
\label{condition_for_alla}
\end{equation}
Terms appearing in Eq.(\ref{condition_for_alla}) are obtained as
\begin{eqnarray}
\sum_i (\bm I)_{i \ell} &=& 1,\nonumber \\
\sum_i \left( \bm A \right)_{i\ell} &=& \alpha\sum_{k}\frac{k_\ell}{k}P(k|k_\ell)=\frac{\alpha k_\ell}{\langle k\rangle},\nonumber \\
\sum_i \left(\bm A^2\right)_{i\ell} &=& \alpha^2\sum_{k_2}\frac{k_1}{k_2}P(k_2|k_1)\sum_{k_1}\frac{k_\ell}{k_1}P(k_1|k_\ell)\nonumber \\
&=&\frac{\alpha^2 k_\ell}{\langle k\rangle},\nonumber
\end{eqnarray}
and we obtain
\begin{eqnarray}
\sum_i (\bm I+\bm A+\bm A^2+\cdots)_{i\ell}=\nonumber \\
1+\sum_{n=1}^{\infty}\frac{k_\ell}{\langle k\rangle}\alpha^n=1+\frac{k_\ell}{\langle k\rangle}\frac{\alpha}{1-\alpha}.
\end{eqnarray}
By replacing the average over individual nodes $\frac{1}{N} \sum_\ell$ in Eq.(\ref{condition_for_all}) by the average over the distribution of degrees $\sum_{k_\ell} P(k_\ell)$, we obtain
\begin{eqnarray}
\frac{1}{N}\sum_\ell \left( \sum_i C_{i\ell} \right)^2 &=& \sum_{k} P(k) \left( 1+\frac{k}{\langle k\rangle}\frac{\alpha}{1-\alpha} \right)^2 \nonumber \\
&=&\frac{\langle k^2\rangle}{\langle k\rangle^2}\left(\frac{\alpha}{1-\alpha}\right)^2+\frac{1+\alpha}{1-\alpha}
\end{eqnarray}
The critical value $\alpha_c$ for the endogenous fluctuation can be obtained by solving 
\begin{equation}
\frac{\langle k^2\rangle}{\langle k\rangle^2}\left(\frac{\alpha}{1-\alpha}\right)^2+\frac{1+\alpha}{1-\alpha} = 2.
\end{equation}
The solution of the above equation $\alpha_c$ is identical to the critical interaction of a one-dimensional self-exciting process $1-1/\sqrt{2}$ if all nodes have identical degrees, $\langle k^2\rangle = \langle k\rangle^2$. However, the critical interaction $\alpha_c$ decreases with the degree dispersion, $\langle k^2\rangle/\langle k\rangle^2$ (Fig.~\ref{fig01}). The scale-free networks characterized by the degree distribution $P(k) \propto k^{-\gamma}$ with $2<\gamma<3$, which are known to be prevalent among social networks~\cite{Kwak10}, exhibit divergence in the degree dispersion $\langle k^2\rangle$, while the mean degree is finite. Thus, the scale-free networks may exhibit endogenous fluctuation with infinitesimal interaction under the input regulation condition.

\begin{figure}[!ht]
\centering
\includegraphics[width=3.4in]{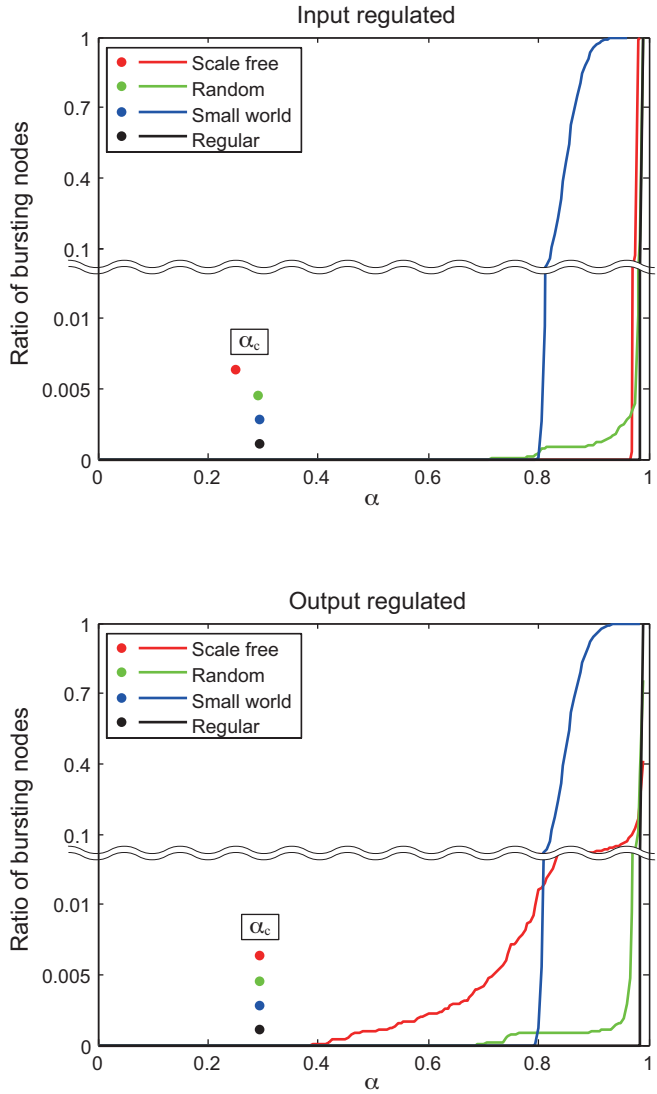}
\caption{Critical interactions $\alpha_c$ for two types of asymmetric interactions computed for uniform, random, small-world, and scale-free networks ($N=10000$). (a) Input regulated condition. (b) Output regulated condition. The fractions of nodes that exhibit fluctuation in individual event series are also displayed.}
\label{figcritical}
\end{figure}

Figure~\ref{figcritical}(a) demonstrates the critical interactions $\alpha_c$ computed for uniform, random, small-world, and scale-free networks of $N=10000$. In this small system, $\alpha_c$ of the scale-free network is still close to that of the uniform connections, though it vanishes in the limit of infinite dispersion. The fractions of nodes that exhibit fluctuation in individual event series are also displayed in Fig.~\ref{figcritical}(a). It should be noted that most nodes do not exhibit fluctuation even though the summed events exhibit bursting. Among three kinds of networks, a small-world network tends to have fluctuating nodes with the smaller interactions than others.

\begin{figure}[!ht]
\centering
\includegraphics[width=3.4in]{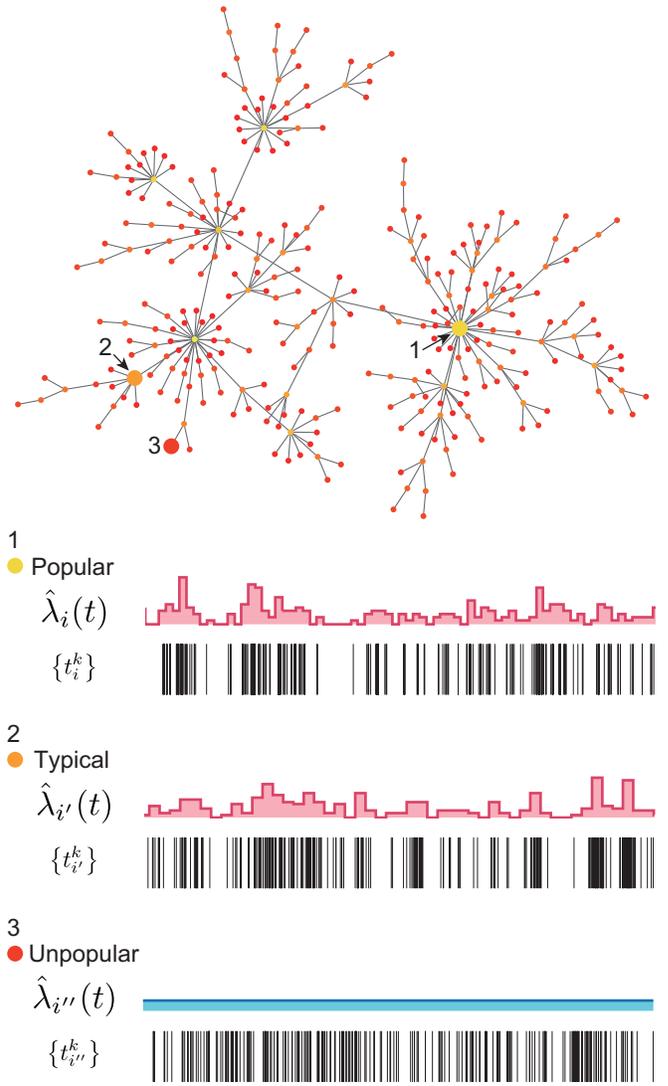}
\caption{Event series sampled from popular, typical, and unpopular nodes. A scale free network of size $n=300$, with the excitability $\alpha=0.7$. The degrees of the (1) popular, (2) typical, and (3) unpopular nodes sampled here are 28, 7, and 1, respectively.}  
\label{figevents}
\end{figure}

\subsection{Output regulation condition}
The condition that the entire output to each node is regulated is represented in such a way that the influence in each interaction is scaled inversely proportional to the degree of $\ell$th node (the sender),
\begin{equation}
\alpha_{i,\ell} = \alpha / k_\ell.
\end{equation}
In this case, the terms appearing in Eq.(\ref{condition_for_alla}) are obtained as
\begin{eqnarray}
\sum_i (\bm I)_{i \ell} &=& 1,\nonumber \\
\sum_i \left( \bm A \right)_{i\ell} &=& \alpha,\nonumber \\
\sum_i \left(\bm A^2\right)_{i\ell} &=& \alpha^2, \nonumber
\end{eqnarray}
and we have
\begin{equation}
\sum_i (\bm I+\bm A+\bm A^2+\cdots)_{i\ell}=\frac{1}{1-\alpha}.
\end{equation}
Accordingly, the condition for the fluctuation being visible is
\begin{equation}
\frac{1}{N}\sum_{i,j}\bm C\bm C^T=\frac{1}{\left(1-\alpha\right)^2}>2,
\end{equation}
and we have $\alpha_c=1-1/\sqrt{2} \approx 0.3$. 

Figure~\ref{figcritical}(b) demonstrates that the critical interactions $\alpha_c$ is identical among three kinds of network topology in the output regulated condition. In the scale-free network, a few nodes start to exhibit fluctuation even when $\alpha$ is close to $\alpha_c$. However, it requires the largest $\alpha$ to make a majority of nodes fluctuate in the scale-free network, in comparison with other networks.

\section{Discussion}

In this study, we demonstrated that the self-exciting point process undergoes a transition through which the rate fluctuation can be detected. In particular, we compared two types of asymmetric interactions in which the total strength of influence is regulated at either input or output. In the input regulation condition, the critical interaction strength decreases with the dispersion of the degrees or the number of connections at individual nodes. The endogenous fluctuation may appear with infinitesimal interaction if the degree dispersion diverges. Contrastingly, in the output regulation condition, the critical interaction strength is constant, identical to that of the one-dimensional Hawkes process and is independent of the degree dispersion. In this way, the endogenous bursting induced by interactions among nodes depends not only on the topology of connections, but also on the manner in which each node is interacting.

\appendix

Here we derive the critical condition for an optimal histogram to be constant, in the case that the bin size of a histogram is selected upon the principle of minimizing the MISE between the histogram and the underlying rate, according to Ref.\cite{Koyama04,shintani12}. 
\begin{figure}[!ht]
\centering
\includegraphics[width=3.4in]{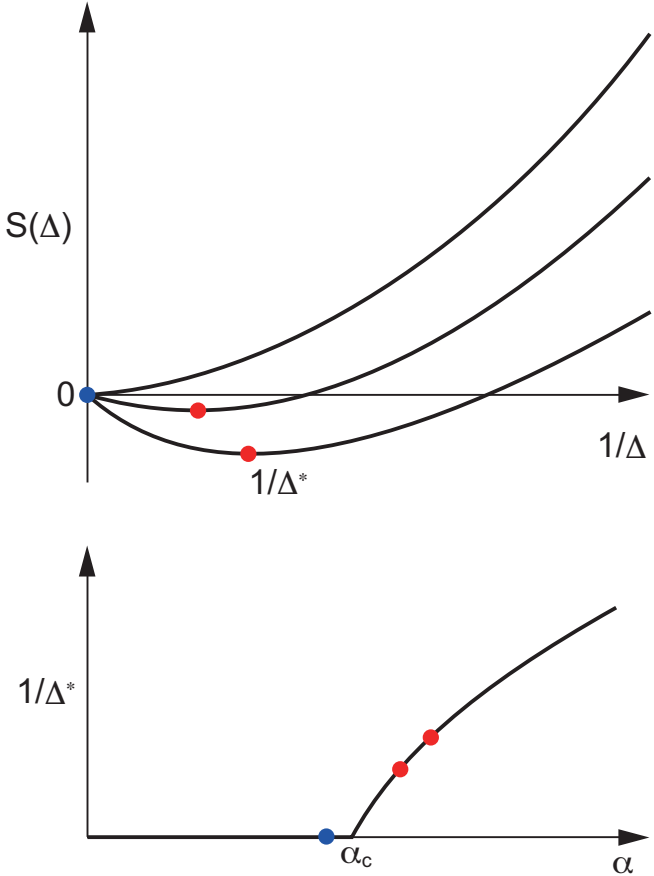}
\caption{Transition in the bin size of MISE-optimal histograms. (top) The MISE $S(\Delta)$ plotted against the inverse of the binsize, $1/\Delta$ for various excitation $\alpha$. (bottom) The inverse optimal binsize, $1/\Delta^*$, plotted against the excitation level $\alpha$.}
\label{estimationtransition}
\end{figure}

The MISE between the underlying rate $\lambda(t)$ and the histogram $\hat{\lambda}(t)$ is given in Eq.(\ref{mise}). In each bin of size $\Delta$, the histogram $\hat \lambda(t)$ is a constant whose height is the number of events $K$ divided by the bin size $\Delta$. Thus the MISE is transformed as
\begin{equation}
S=\left\langle \frac{1}{\Delta} \int_0^{\Delta} \left( \lambda^2(t) -\frac{2 K}{\Delta} \lambda(t) + \frac{K^2}{\Delta^2} \right) dt \right\rangle.
\label{mise1}
\end{equation}
The expected number of events in each interval is given by integrating the underlying rate: $\left\langle K \right\rangle=\int_0^{\Delta} \lambda(t) dt$. Because events are independently drawn, the Poisson relation holds: $\left\langle K^2 \right\rangle=\left\langle K \right\rangle^2+\left\langle K \right\rangle$. Inserting these relations into Eq.(\ref{mise1}), we have
\begin{equation}
S=\phi(0)+\frac{\left\langle \lambda \right\rangle}{\Delta}-\frac{1}{\Delta^2}\int_0^{\Delta}dt\int_{-t}^t \phi(s) ds,
\end{equation}
where $\phi(s) \equiv \left\langle \lambda(t+s)\lambda(t) \right\rangle-\left\langle \lambda \right\rangle^2$ is the correlation of the rate fluctuation, or 
\begin{equation}
\phi(s) = \left\langle \delta \lambda(t+s) \delta \lambda(t) \right\rangle,
\end{equation}
where $\delta \lambda(t) \equiv \lambda(t) - \left\langle \lambda \right\rangle,$ is the temporal fluctuation of the rate.

For a homogeneous Poisson process for which $\phi(s)=0$, the MISE is a monotonically decreasing function, $S = \left\langle \lambda \right\rangle / \Delta$, and therefore, the optimal bin size diverges. 

By contrast, the MISE of inhomogeneous point processes may have a minimum at some finite $\Delta$. Based on the second order transition in which the minimum appears continuously from the infinite $\Delta$ or infinitesimal $1/\Delta$ (Fig.~\ref{estimationtransition}), the condition for the transition is given as
\begin{equation}
\left. \frac{dS}{d(1/\Delta)}\right|_{\Delta=\infty} < 0.
\end{equation}
This can be summed up as a condition of the rate fluctuation, given in the inequality (\ref{estimation}), on condition that $\int_0^{\infty}s \phi(s) ds$ is finite. 

\section*{ACKNOWLEDGMENTS}

This study was supported partly by Grants-in-Aid for Scientific Research to SS from the MEXT Japan (25115718, 26280007), and by JST, CREST.

\bibliographystyle{IEEEtran}
\bibliography{IEEEabrv,reference}

\end{document}